\documentclass{egpubl}
\usepackage{EGUKCGVC2021_XX}
 
\WsSubmission      

\usepackage[T1]{fontenc}
\usepackage{dfadobe}  

\usepackage{cite}  
\BibtexOrBiblatex
\electronicVersion
\PrintedOrElectronic
\ifpdf \usepackage[pdftex]{graphicx} \pdfcompresslevel=9
\else \usepackage[dvips]{graphicx} \fi

\usepackage{egweblnk} 

\title[Automating Visualization Quality Assessment]%
      {Automating Visualization Quality Assessment: \\ a Case Study in Higher Education}

\author[N.S. Holliman]
{\parbox{\textwidth}{\centering N.\,S. Holliman\thanks{Fellow of The Alan Turing Institute, London}$^{1}$ } \\
{\parbox{\textwidth}{\centering $^1$School of Computing, Newcastle University, United Kingdom} } }

\begin{document}


\maketitle
\begin{abstract}
We present a case study in the use of machine+human mixed intelligence for visualization quality assessment, applying automated visualization quality metrics to support the human assessment of data visualizations produced as coursework by students taking higher education courses. A set of image informatics algorithms including edge congestion, visual saliency and colour analysis generate  machine analysis of student visualizations. The insight from the image informatics outputs has proved helpful for the marker in assessing the work and is also provided to the students as part of a written report on their work.
Student and external reviewer comments suggest that the addition of the image informatics outputs to the standard feedback document was a positive step. 
We review the ethical challenges of working with assessment data and of automating assessment processes.


\end{abstract}  
\section{Introduction}

Over the last two academic years, we have been evaluating the use of automatic methods for assessing the quality of visualizations submitted as coursework by students on our Data Science MSc programme. This is part of a strand of work on visualization quality metrics (VizQM) that we are developing in order to both help inform visualization developers about the quality of their visualizations and in the longer term as we look towards automating the process of visualization production itself.

\section{Background}

The automatic assessment of computer science coding assignments has a history going back to the earliest days of teaching programming~\cite{Hollingsworth1960}. Since then, a wide range of methods have been developed to automate the process of testing and marking student programming assignments~\cite{Wilcox2015}.

The evaluation of the effectiveness of visualizations using experimental methods has a shorter history; one review in 2004~\cite{plaisant2004} identified four approaches: controlled experiments to test design choices, usability evaluations (UX), tool comparisons and case studies in realistic settings. Concluding that "..visualization research must understand the principles that will help the field cross the chasm to wider success".

Recently the idea that visualizations could be automatically assessed using algorithmic quality metrics has been growing in the literature. This has included using VizQM as a fitness function in a production optimisation loop ~\cite{House2006,Holliman2015,micallef2017} and as a tool for measuring the effectiveness of the insight a visualization can provide~\cite{North2006}.  

In our teaching, we define a set of visualization principles as learning objectives and then in summative exercises, seek to test student understanding of these principles against a set of marking and feedback criteria. In human marking, we judge the achievement against the criteria and explicitly mark the work and provide written feedback. In this case study, we asked the question to what extent can automatic image informatics algorithms support the process of generating marks and feedback?

\subsection{The Objective of Automating Assessments}

In reviewing the role of automation in undergraduate computer science education Wilcox~\cite{Wilcox2015} suggests two objectives for automatic assessment :
\begin{itemize}
    \item Does the proposed automation contribute to or detract from student learning?
    \item Do the benefits outweigh the costs of automation?
\end{itemize} 

Here our goal is to evaluate a mixed intelligence approach, rather than full automation. This differs from some coding assignments where data output comparisons and code checking tools can automatically generate proposed marks, although there are questions about the reliability of these in all cases \cite{wrenn2018}.

\subsection{Visual Quality Metrics}
There is a wide range of literature in visual quality metrics for image~\cite{Zhai2020} and video~\cite{Maia2015} coding standards which seek to measure subjectively and objectively the effect of image compression algorithms on perceived image quality. In visualization, similar visual quality metrics exist but are less well developed and have a number of dimensions that are distinct to the purpose of visualization in conveying knowledge~\cite{carpendale2008}.

A recent review of visualization quality metrics~\cite{Behrisch_VizQM_2018} sets out to classify metrics as related to low, mid or high-level perception. The authors argue that extant clutter based approaches to quality should be superseded by measures of visual pattern retrieval. While the review is thorough and comprehensive, it could have gone further in linking to interdisciplinary work in experimental psychology and ethical philosophy where validated models of perception and cognition are being proposed~\cite{burns2020}. 

One example of work that crosses over from experimental psychology to visualization is this study of the capabilities and limits of peripheral vision~\cite{rosenholtz2016}. The author considers how better models of peripheral vision can predict the perception of the gist of a scene from a glance and which in turn might be used to predict which visualization design approaches would enable quicker knowledge perception.

An example of a review of the emerging field of the philosophy of visualization is presented in~\cite{Engebretsen2020}. This brings into focus what we mean by visualization socially and politically and how insight is constructed in visual presentations of data. When visualizations can engage people globally in a matter of hours, it becomes as important as ever to understand the influence visualizations can have on human cognition. However, algorithmic metrics for these aspects of the understanding of visualizations are not widely studied; for example, we are currently unaware of any metrics for the sentiment analysis of a data visualization. 

\subsection{Image Metrics for Visualization Quality}

While there are very many image metrics that could be applied to measure visualization quality we looked for a set of metrics that could be applied to student work relatively quickly and easily. To support these we used the Python Imaging Library~\cite{pillow} for basic image handling functions including image file reading, writing and resizing.

The OpenCV library~\cite{opencv2008} provides a number of useful tools including a fast low-resolution saliency prediction method. While we did integrate this into our toolset we will concentrate in this discussion on the fine detail saliency metric described below. 

Another saliency metric we have tested for use in the metrics is the DeepGaze II algorithm ~\cite{Kummerer_2017_ICCV}. This is a deep convolutional neural network trained to predict salient features in a still image, it ranks highly in the MIT/Tuebingen benchmark~\cite{mit-tuebingen-saliency-benchmark} when tested against gold standard eye-tracking measures of visual salience. However, as we have not yet used the DeepGaze algorithm over repeated cohorts we do not report in detail on it here.

\begin{figure}[htb]
  \centering
  \includegraphics[width=.95\linewidth]{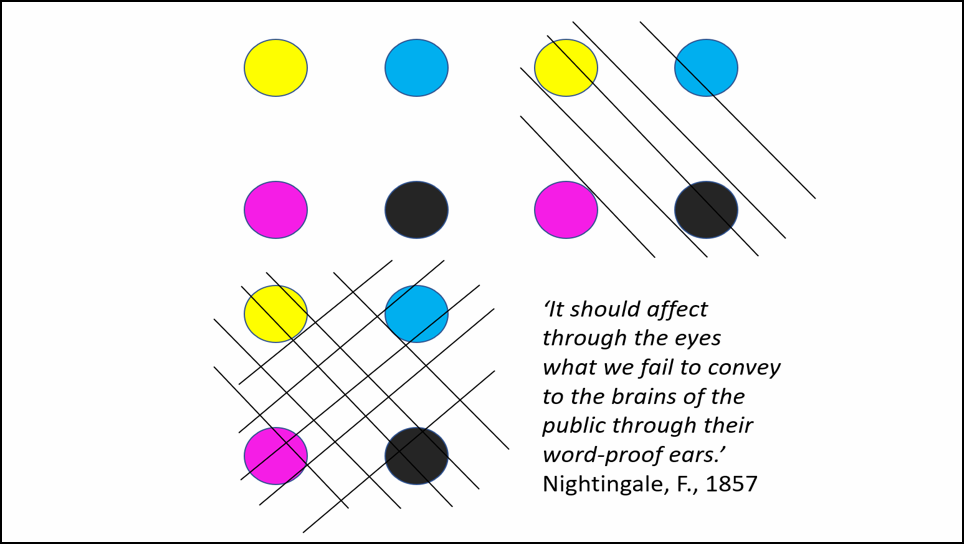}
  \caption{\label{fig:testImage}
           We use this image to illustrate the effect of the metrics we have chosen to use, it contains basic colour, clutter and text items.}
\end{figure}

The image metrics we discuss in this article all come from the Aalto Interface Metrics ({AIM}) system~\cite{Oulasvirta:2018} repository. The AIM system is available as an online tool that can apply a battery of image tests to web pages and graphical user interfaces. After testing we decided the following metrics were the most appropriate for the needs of a visualization assessment toolkit. We use the test image in Figure~\ref{fig:testImage} to illustrate the basic operation of the metrics.

\subsubsection{Edge Congestion}
\begin{figure}[htb]
  \centering
  \includegraphics[width=.95\linewidth]{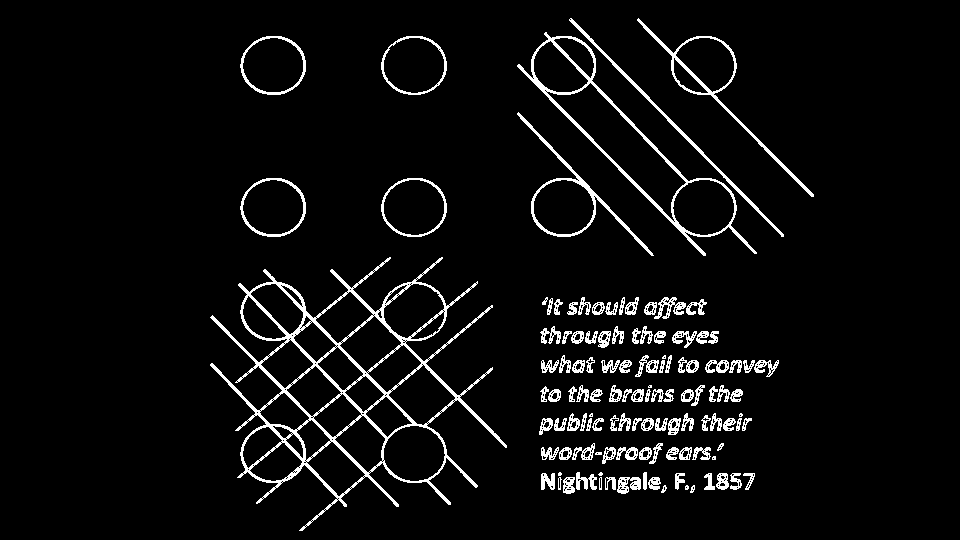}
  \caption{\label{fig:testCongestion}
           The output from the edge congestion metric for the test image, white pixels are flagged as congested and potential areas of visual complexity in the image.}
\end{figure}

Edge congestion as a metric has been demonstrated to correlate with the perceived complexity of a GUI image and negatively correlate with its perceived aesthetic quality \cite{miniukovich2014}. We adopted this edge congestion method as it accounts for colour in its calculation, the python code is available from the AIM project github pages~\cite{Oulasvirta:2018} and is based on an original method described in ~\cite{rosenholtz2007}. The output for the test image is shown in Figure~\ref{fig:testCongestion}. Note the threshold value for the distance in pixels between two edges to be labelled as congested edge is set as 4 pixels since the original suggested in \cite{miniukovich2014} of 20 pixels proved to be too large. 

\subsubsection{Saliency}
Saliency refers to the visual conspicuousness of regions of the visualization. It tries to predict which areas of the image will attract attention when glancing 
at the visualization or when searching for information. Fine detail saliency \cite{itti2000} computes detailed saliency of shapes and areas, where brighter pixels are more salient. The machine vision algorithm we used here accounts for image colour in its saliency calculation and we use an implementation of the Itti-Koch algorithm by Akisato Kimura~\cite{Oulasvirta:2018}.

\begin{figure}[htb]
  \centering
  \includegraphics[width=.95\linewidth]{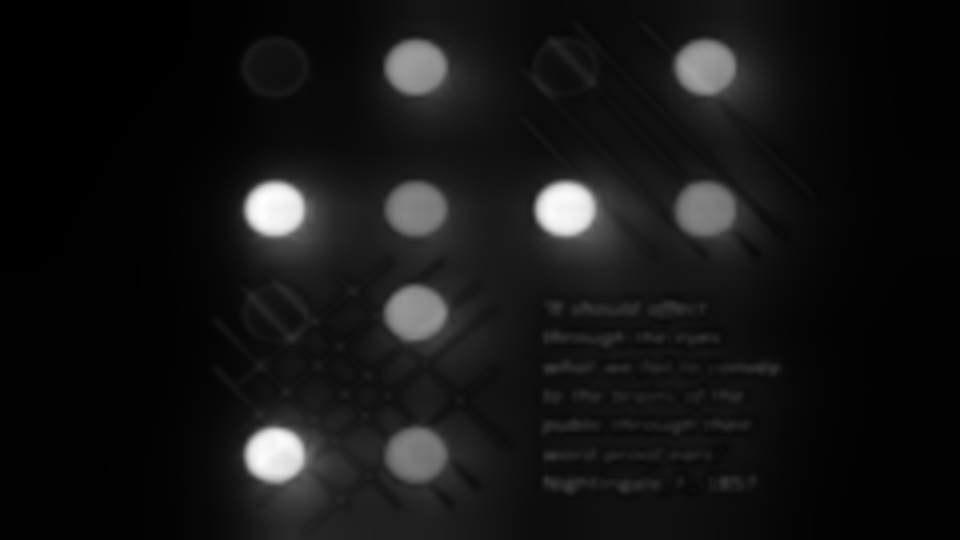}
  \caption{\label{fig:testSaliency}
           The output from the fine detail saliency metric for the test image, white pixels are flagged as high saliency and should correspond to the key data in a visualization.}
\end{figure}

One simple, subjective way to evaluate the saliency output is to compare where your eyes are drawn to in the original image with the predictions from the saliency calculations in Figure~\ref{fig:testSaliency}. The saliency algorithm here predicts the coloured discs are more salient than the text and much more salient than the clutter lines. A possible weakness of this particular algorithm is also illustrated, it tends to rate yellow areas as having low saliency.

\subsubsection{Colour Metrics}
\begin{figure}[htb]
  \centering
  \includegraphics[width=.95\linewidth]{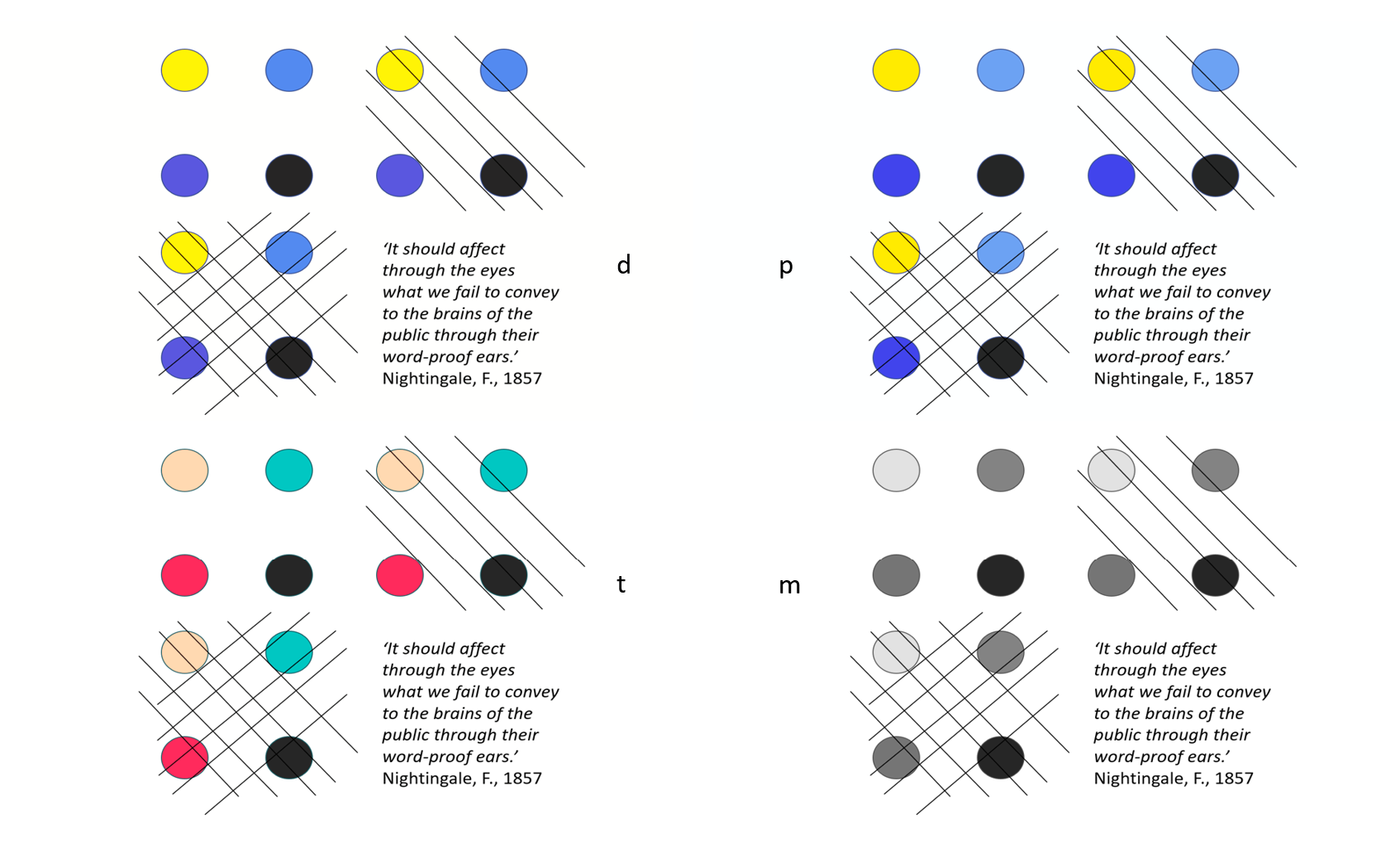}
  \caption{\label{fig:testColour}
           A colour vision analysis of the test image in four ways; simulating deuteranomaly, protanomaly and tritanomaly colour deficiencies, and grey-scale viewing of the test image. }
\end{figure}

Colour and colour vision are complex subjects, but perhaps one of the best understood in terms of quality metrics. A reasonable percentage of people have some level of colour deficiency therefore you might want to check your own colour vision before reviewing the outputs in Figure~\ref{fig:testColour}.

The three images labelled \textit{ d, p} and \textit{t} in Figure~\ref{fig:testColour} are simulations of these three colour deficiencies:
\begin{itemize} 
\item Deuteranomaly: A red-green deficiency lacking more in green cones.
\item Protanomaly: A red-green deficiency lacking more in red cones.
\item Tritanomaly: A rare blue-yellow deficiency lacking in blue cones.
\end{itemize} 
These are calculated by the colour vision deficiency simulation method ~\cite{machado2009} using Python code shared by the {AIM} project~\cite{Oulasvirta:2018}. 

The fourth image in Figure~\ref{fig:testColour} labelled \textit{m} is a monochrome grey-scale simulation to replicate the result of either printing or faxing the visualization. This is important to review as frequently visuals are presented in printed form and key information should ideally still be preserved. The monochrome image is calculated using the {OpenCV} {COLOR\_RGB2GRAY} method \cite{opencv2008}.

We also calculate two subjective quality metrics relating to colour, again with Python code shared via the  AIM project. 

\begin{figure*}[htb]
  \centering
  \includegraphics[width=.95\linewidth]{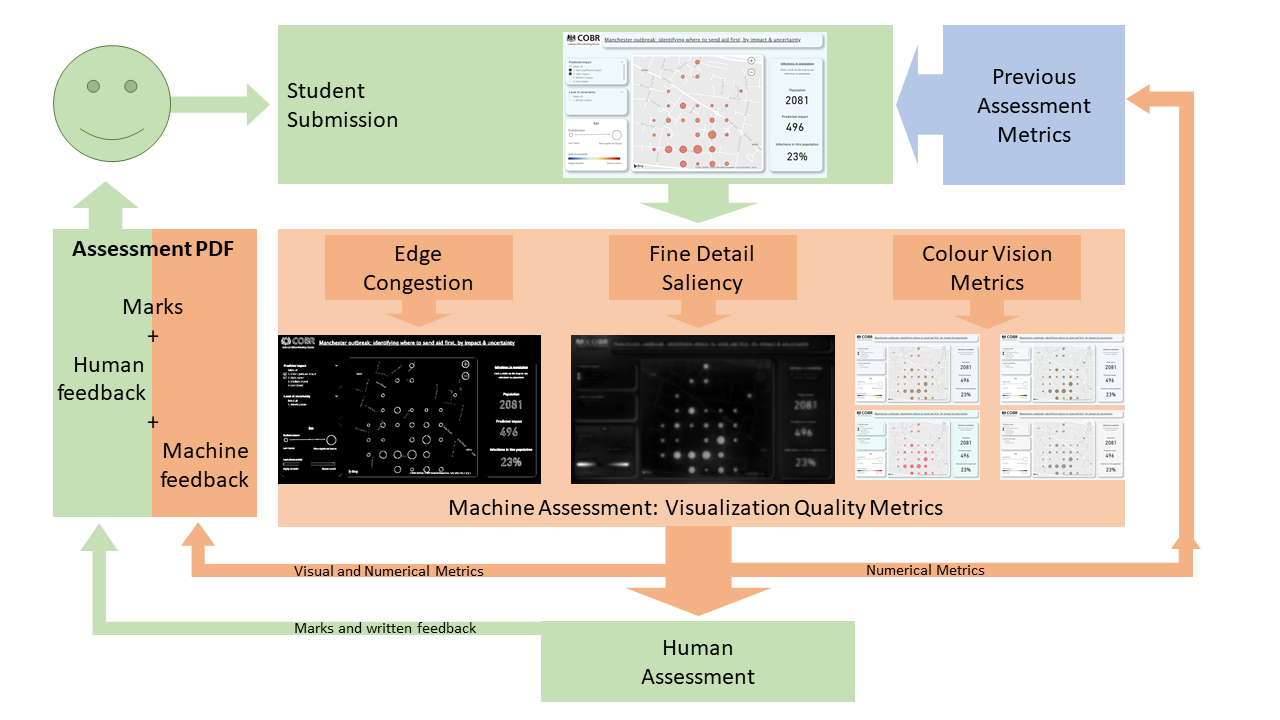}
  \caption{\label{fig:process}
           Our mixed intelligence process generates feedback and marks by combining human and machine analysis of a student's visualization. Each snapshot of a submission is automatically analysed and the results fed to the human marker who uses them to inform marking ad feedback, the combined machine and human assessment is written into a PDF file that is then returned to the student.}
\end{figure*}

The WAVE metric~\cite{palmer2010_WAVE} calculates a colour preference score which is underpinned by experimental data from surveys of people's colour preferences. Higher values of this metric suggest viewers prefer to see the colours in the visualization. 

The second numerical measure is the Hasler-Susstrunk~\cite{hasler2003} metric which correlates with judgements of colourfulness. Higher values suggest more people prefer the colours used in the visualization and this metric been shown to correlate with aesthetic impression in photographs.

\subsection{Display and Viewing Environment}

A factor that is rigorously controlled in most human perception experiments is the display screen characteristics and the viewing environment. Perhaps because of the difficulty of generally doing this in practice very few visualization metrics or experiments collecting data fully consider this.  While this is a pragmatic solution, it is important to note that viewing distance, display size and pixel resolution as well as the screen luminance range and gamma response curve all make substantial differences to the visibility of fine detail and colour gradations. In the following, we don't consider this further but do flag it as a critical area for future consideration. 

\section{Visualization Assessment Automation Case Study}

\begin{figure}[htb]
  \centering
  \includegraphics[width=.95\linewidth]{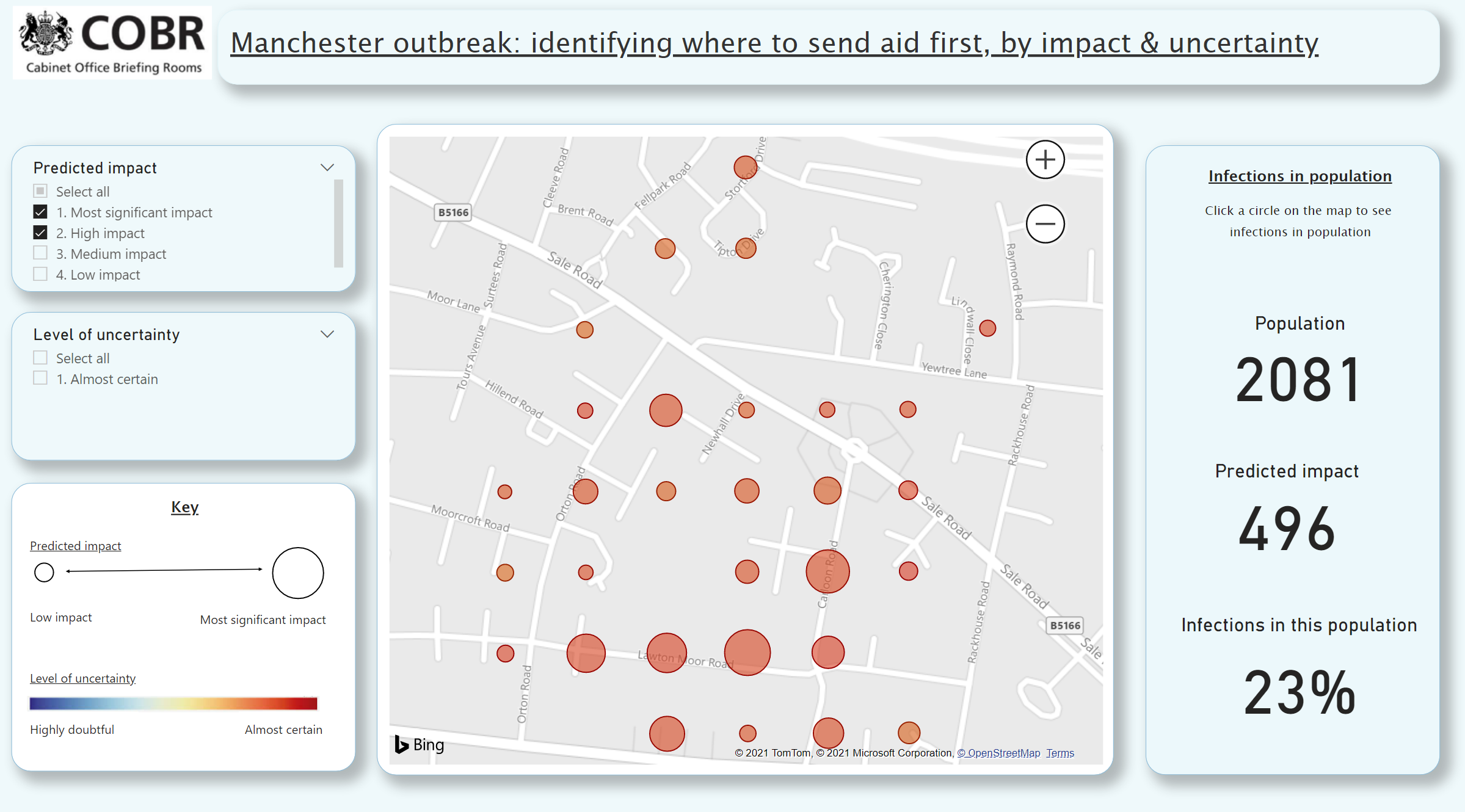}

  \caption{\label{fig:output01}
           This is a screenshot from one submitted student assessment. It shows a geographic pattern of a simulated disease outbreak where the glyphs represent both the impact (size) and uncertainty of impact (colour). The goal is to allow a user to highlight the most certain, highest-impact areas to send aid to first.}
\end{figure}

The case studies we now work through in detail are both examples of the assessment of a Masters level assignment in a Data Science course. The students are required to produce a visualization of the output data from a cell-based simulation of an infectious agent spreading across a city. The scenario requires the presentation of simulation output data predicting the likely number of infections in each cell and the uncertainty of these estimates as shown in Figure~\ref{fig:output01}.

As this course was already running with human-only assessment a set of marking and assessment criteria were available.
The learning outcomes for the assignment and the visual elements that relate to these are summarised in Table~\ref{fig:appendix}.

\subsection{Assessment Process Overview}

An overview of the mixed intelligence process that we use is shown in Figure~\ref{fig:process}. From top left a student submission flows through the VisQM analysers, and the machine assessment output is fed back to the human marker. This consists of both an image from each VisQM metric and a calculated ranking against all previous visualizations the system has seen. 

The human assessor then uses the machine assessment information to inform the marks given for the work and to support the written feedback on the submission. All this information including marks, human feedback and machine feedback is then combined into an eight-page PDF file for return to the student. 

\section{VizQM Results Walk-through}

We walk-through eight of the ten VizQM that we apply to each visualization to highlight what is calculated and how each contributes to the human marking and feedback. The two remaining VizQM not discussed for space reasons are low-resolution saliency and simple edge detection. The first set of results discussed here relate to the student work shown in Figure~\ref{fig:output01}.

\subsection{Edge Congestion}

\begin{figure}[htb]
  \centering
  \includegraphics[width=.95\linewidth]{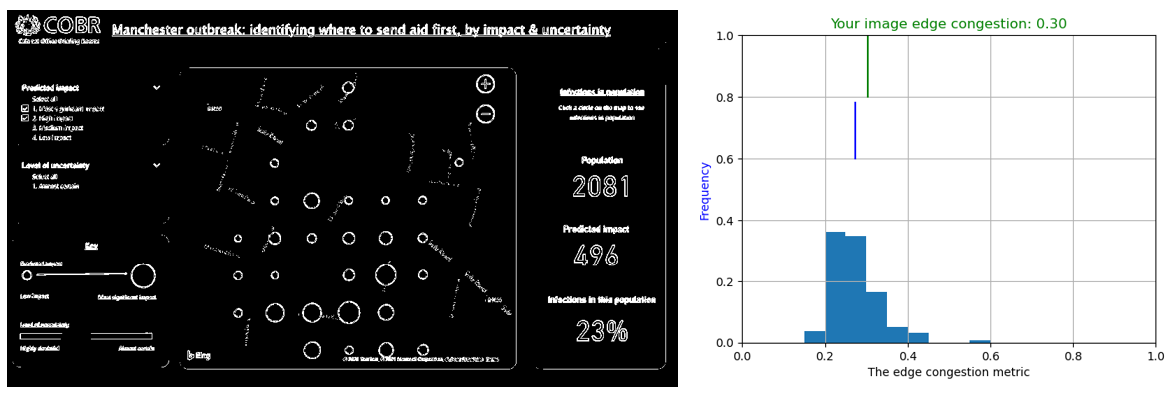}

  \caption{\label{fig:edgeResults}
           The edge congestion output and a ranking of this output compared to all other visualizations the system has seen to date using the edge congestion score, $S_{ec}$, shown as the green vertical line.}
\end{figure}

The edge congestion metric highlights edges that are in the visualization which are crowded based on a distance measure and also accounts for colour. The image output for the visualization in Figure~\ref{fig:output01} is shown in Figure~\ref{fig:edgeResults}.

As part of the process of calculating the edge congestion image, all edges in the image are calculated. The allows an edge congestion score $S_{ec}$ to be calculated as the ratio of pixels in congested edges to the ratio of all edge pixels in the image.

This image and the $S_{ec}$ ranking can be used by the human marker as a guide to the clarity of the image, for example, if there is a high ranking it suggests an unusually cluttered image. In addition, the layout of key data items is visible and this contributes to written feedback on the use of gestalt design methods. For example,  highlighting if there is an appropriate use of space or enclosure between related data elements. 

\subsection{Saliency}

\begin{figure}[htb]
  \centering
  \includegraphics[width=.95\linewidth]{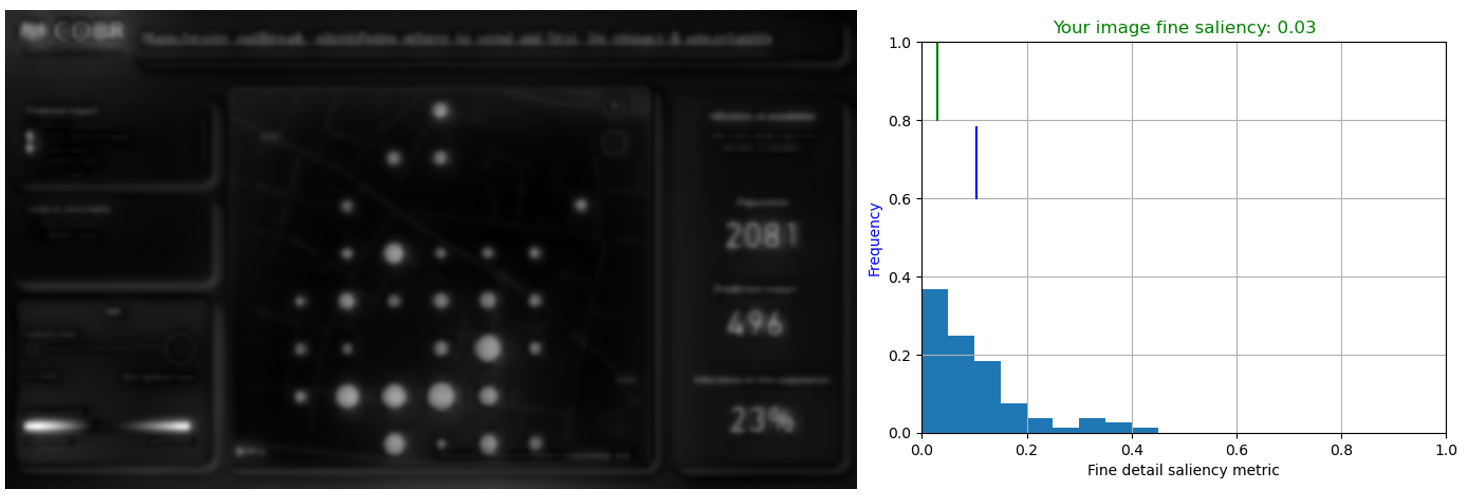}

  \caption{\label{fig:salicResults}
           The saliency calculation output and a ranking of this output compared to all other visualizations the system has seen to date using the saliency score, $S_{sy}$, shown as the green vertical line.}
\end{figure}

The fine detail saliency metric, shown in Figure~\ref{fig:salicResults}, can predict where the eye is drawn to in the visualization. The hope is that the most important aspects of the visualization are highlighted, for example, the data glyphs more so than the title.

The saliency score $S_{sy}$ is calculated as the ratio of salient pixels to all pixels in the image. Salient pixels are calculated using by thresholding the saliency output against a constant threshold value $I_k = 64$, which was obtained heuristically through experimentation.

The saliency image is used by the human assessor to identify whether salient pixels are linked to the key data items in the image. In Figure~\ref{fig:salicResults} this is true, however, the key at the lower left is the most salient item on the screen suggesting that there is room to improve the representation of the key data glyphs. In this case, written feedback could be that using solid filled discs instead of semi-transparent ones would help, as would using a brighter colour.

Experience suggests that a low saliency score, $S_{sy}$, suggests  feedback for the student that contrast in the visualization could be improved to better highlight key items. Saliency can also highlight which edges and enclosures are most visible to support feedback on the enclosure and connectedness gestalt design principles.

\subsection{Colour Metrics}

The machine evaluation of colour for each visualization is slightly different, six algorithms are applied to each visualization. Four of these output only image feedback and two of them only output numerical scores that are ranked against previously seen visualizations.
\begin{figure}[htb]
  \centering
  \includegraphics[width=.95\linewidth]{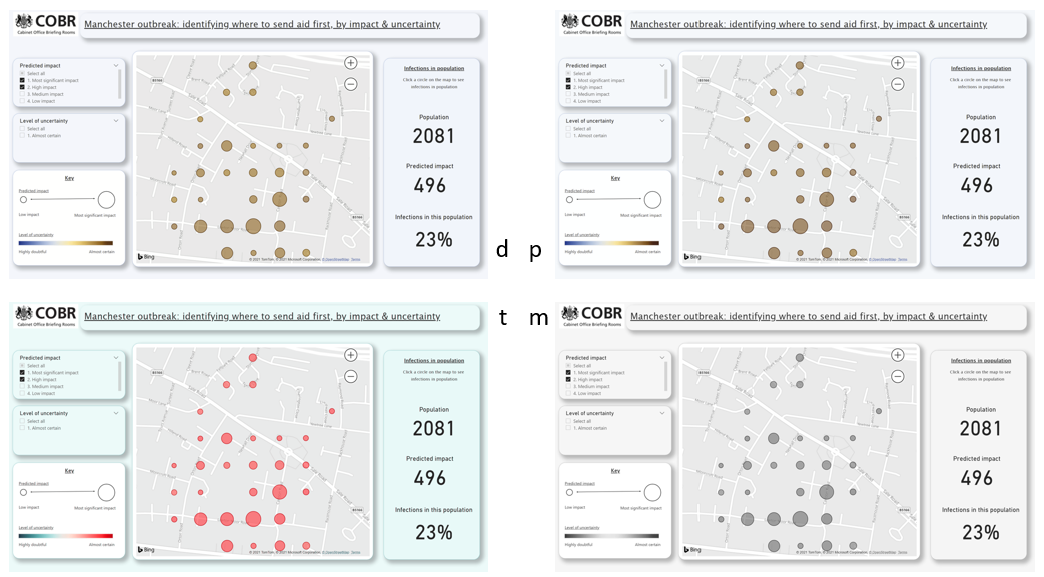}

  \caption{\label{fig:colResults}
           A colour vision analysis of the submitted visualization in four ways; simulating deuteranomaly, protanomaly and tritanomaly colour vision deficiencies, and monochrome grey-scale viewing of the test image.}
\end{figure}

The four images give immediate feedback to students and also support human marking and feedback providing direct evidence of the success of the colour choices in the four conditions tested. Diverging colour scales such as used here can often give problems as they rarely work well in all tested conditions, for example in the monochrome condition simulated here both low and high uncertainty values have the same low grey-scale value.

\begin{figure}[htb]
  \centering
  \includegraphics[width=.95\linewidth]{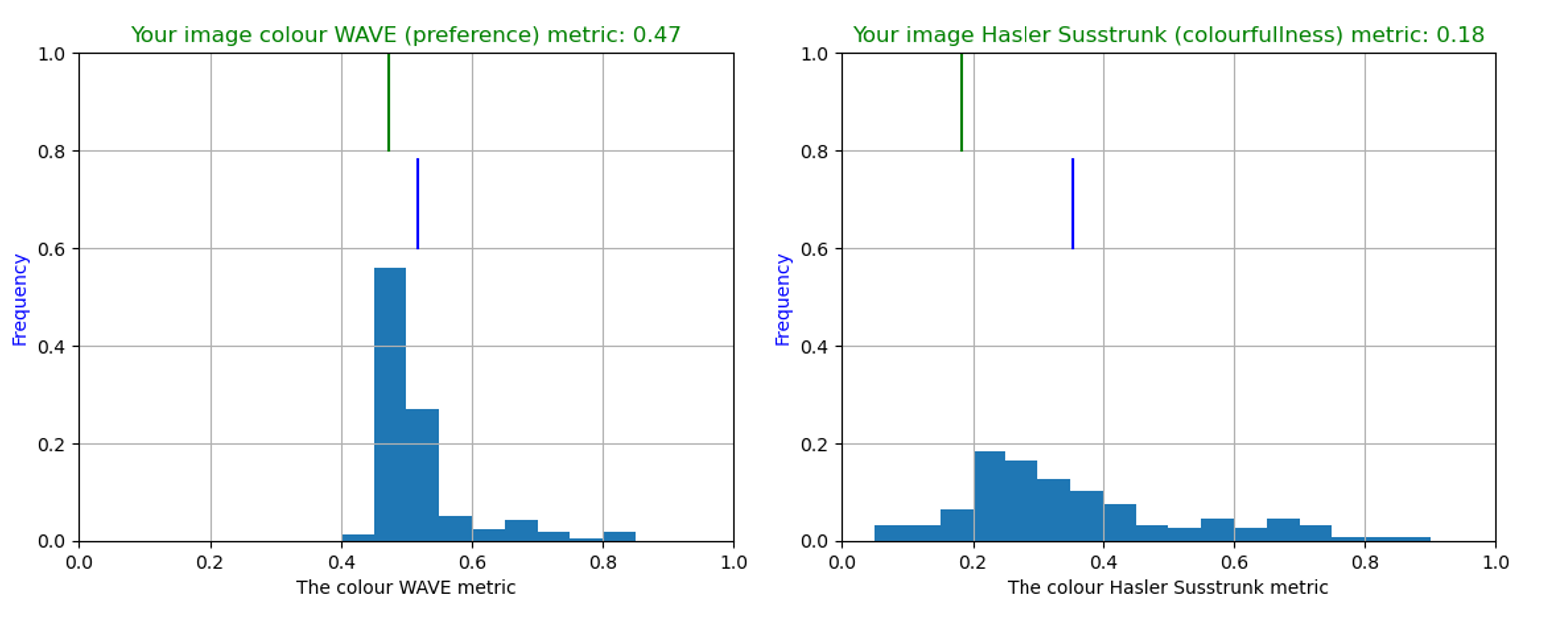}
  \caption{\label{fig:colMetricResults}
           The two numerical metrics, $S_{wv}$ WAVE and $S_{hs}$ Hassler-Susstrunk, are shown as green lines ranked for the submitted visualization against all previously seen visualizations.}
\end{figure}

The $S_{wv}$ WAVE and $S_{hs}$ Hassler-Susstrunk metrics help support feedback and marking on the subjective aspects of the visualization. The $S_{wv}$ metric tends to cluster strongly but some visualizations do pick colours that are particularly high on the preference scale. The $S_{hs}$ metric has a greater spread but higher values really do seem to correlate well with bright and colourful visualizations. The visualization tested here ranks low on colourfulness and this suggests feedback to the student that it will be less memorable, but also perhaps clearer than other more colourful visualizations.

\section{Comparative example}
\begin{figure}[htb]
  \centering
  \includegraphics[width=.95\linewidth]{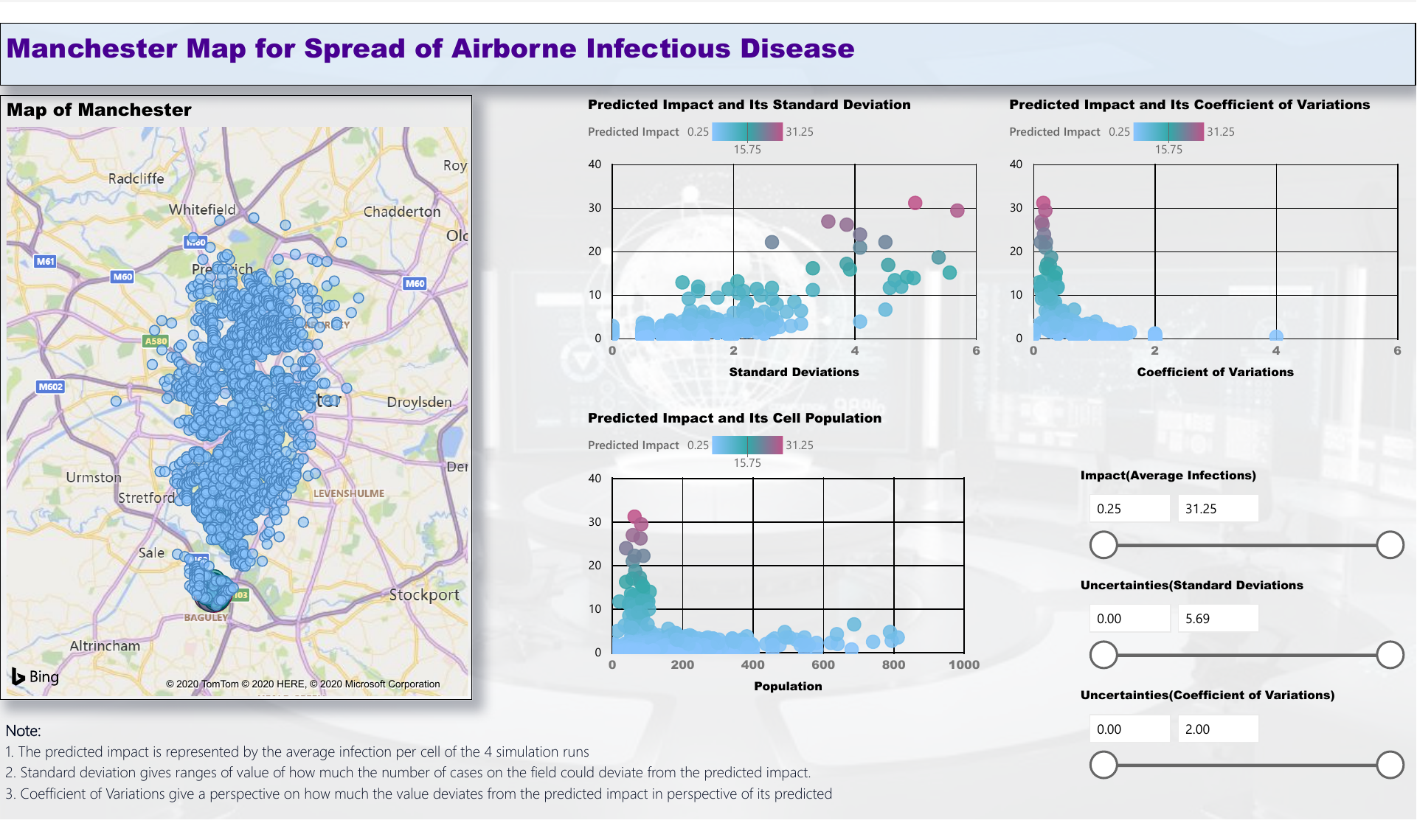}
  \caption{\label{fig:output03}
           This is a screenshot from a second submitted visualization. It shows a geographic pattern of a simulated disease outbreak where the glyphs represent both the impact and uncertainty of impact. The goal is to allow a user to highlight the most certain, highest-impact areas to send aid to first.}
\end{figure}

For comparison we include a second visualization submitted by a student from the same cohort, see Figure~\ref{fig:output03}.

\subsection{Congestion}
\begin{figure}[htb]
  \centering
  \includegraphics[width=.95\linewidth]{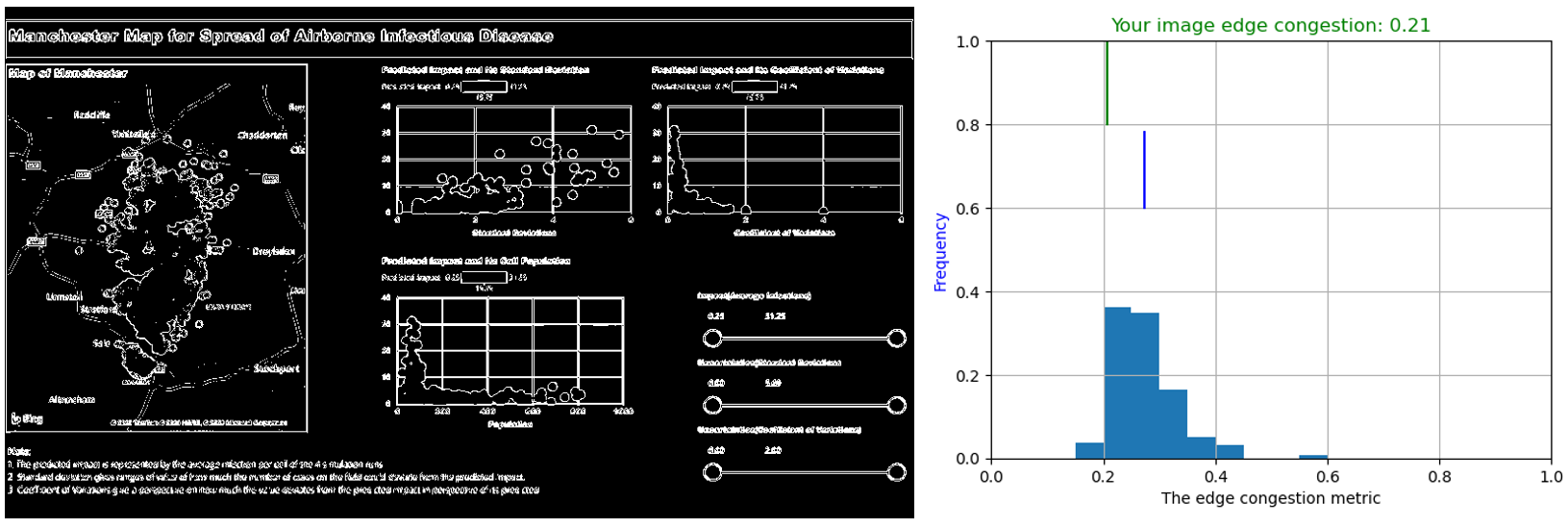}
  \caption{\label{fig:ST03edgeResults}
           The edge congestion output and a ranking compared to all other visualizations the system has seen to date using the edge congestion score, $S_{ec}$, shown as the green vertical line.}
\end{figure}

The edge congestion analysis, Figure~\ref{fig:ST03edgeResults}, suggest this image ranks similarly to most other visuals tested. The image highlights the layout showing how the different elements are grouped with more space separating less related items. It also suggests potential feedback that the gridlines could be removed, reducing non-data ink in the visualization.

\subsection{Saliency}
\begin{figure}[htb]
  \centering
  \includegraphics[width=.95\linewidth]{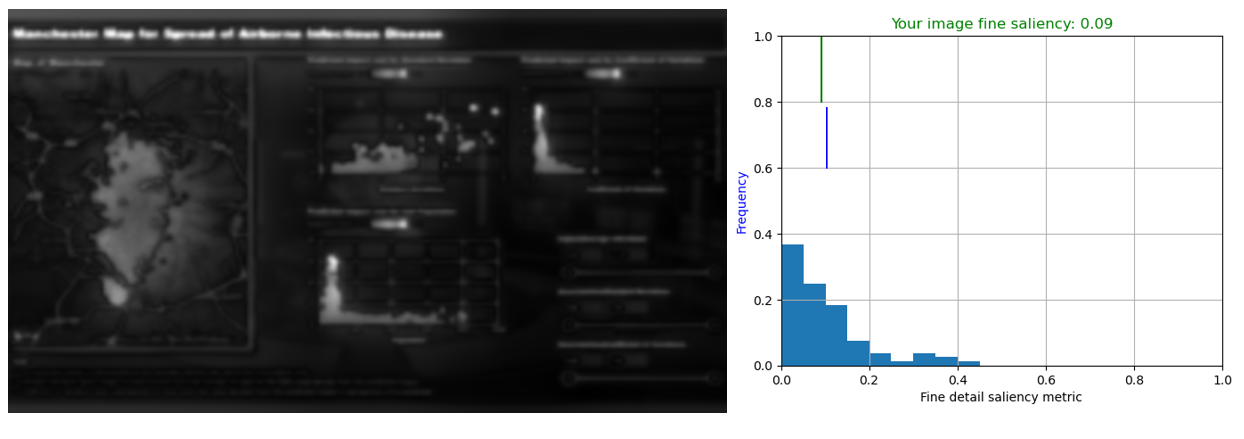}
  \caption{\label{fig:ST03salicResults}
           The saliency calculation output with the saliency score for this visualization, $S_{sy}$, shown as the green vertical line.}
\end{figure}

Ths saliency result, Figure~\ref{fig:ST03salicResults}, suggests that the most salient features in the visualization are the data, other than the title of the page. This suggests weaker contrast in the page title would help draw visual attention to data results.

\subsection{Colour}
\begin{figure}[htb]
  \centering
  \includegraphics[width=.95\linewidth]{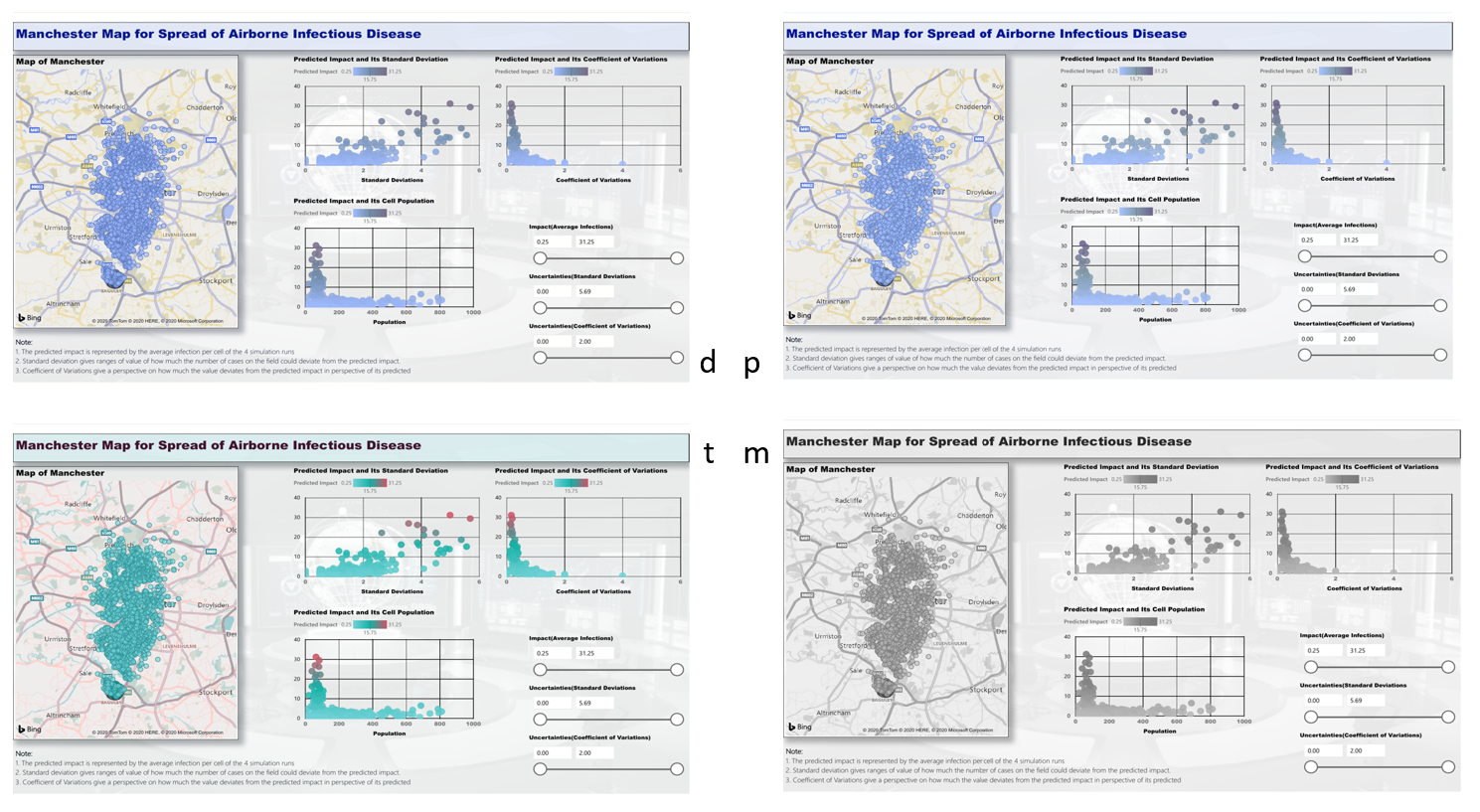}
  \caption{\label{fig:colResults03}
           A colour vision analysis of the submitted visualization in four ways; simulating deuteranomaly, protanomaly and tritanomaly colour vision deficiencies, and monochrome grey-scale viewing of the test image.}
\end{figure}

\begin{figure}[htb]
  \centering
  \includegraphics[width=.95\linewidth]{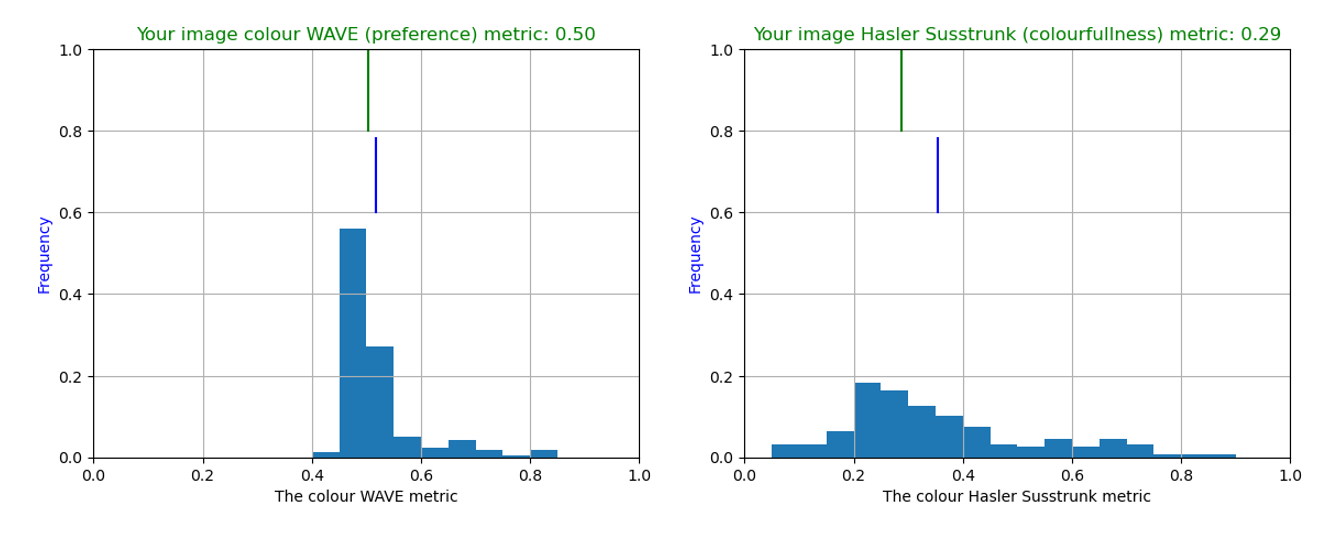}
  \caption{\label{fig:colMetricResults03}
           The two numerical metrics, $S_{wv}$ WAVE and $S_{hs}$ Hassler-Susstrunk, are shown as green lines ranked for the submitted visualization against all previously seen visualizations.}
\end{figure}

The colour vision simulation results, Figure~\ref{fig:colResults03}, suggest that this visual works well across all conditions this is largely due to the almost monochrome scale for values. Only the very highest values are red, and when not visible this shifts to a dark grey-scale value which still works. The two aesthetic metrics, Figure~\ref{fig:colMetricResults03}, rank this visualization close to average on colour preference, but with a largely monochrome colour scheme it ranks a little low on colourfulness.

\section{Ethical Considerations}

In the current complex ethical context, it is worth re-emphasising that the goal of the work here was not to automatically determine student grades for their visualizations. It was instead to generate image-informatics results that inform the human feedback and marking process and additionally provide new types of machine feedback to students. In this mixed intelligence approach, machine plus human, there needs to be consideration of the potential ethical concerns that machine feedback may introduce. 

\subsection{Ensuring Student Confidence}
The UK Office for Statistical Regulation produced a report~\cite{OSR2021} investigating failures of processes involving algorithms in grading {A-levels} in the UK in 2021, this has the following key recommendations for future algorithmic processes used in student assessment:
\begin{itemize}
    \item Be open and trustworthy.\\
    To implement an open process we were careful to include context for the students on how the machine assessment was used in the assessment and referenced all algorithms in the feedback documents. 
    \item Be rigorous and ensure quality throughout. \\
    This article itself is part of our effort to rigorously document this work and to reflect on the quality of the assessment achieved. 
    \item Meet the need and provide public value.\\
    Student feedback to date has been positive and suggests that this addition to the marking process is valued and helps meet a need for more insightful feedback.
\end{itemize}
 It is an open question whether we should provide a version of the machine assessment tool to future students during the production of their coursework. This could help provide feedback during the visualization design process, supporting the work we do in practical laboratories to guide students in their use of visualization tools and techniques.

\subsection{Privacy of Personal Data}
One difficulty in researching new approaches for assessment is the need to conform to data protection regulations~\cite{GDPR2018}, including obtaining student permission to use their work since the output from an assessment of student work is personal data. 

We have taken care to gain permission to use and discuss the visualizations in this report, however, not all students are happy to give permission. This inherently limits how open we can be about the system using real student data and introduces a possible risk of selection bias. For example, higher-achieving students may be a lot happier for us to share results than lower-achieving students. In the future, this could be a motivation for the generation of representative synthetic data sets, a topic of some interest across many areas of data science~\cite{ONS2019}.

\subsection{High-Risk Algorithms}
In April 2021 the European Commission published a draft regulation for harmonised rules on the use of AI systems within the EU~\cite{EC2021} . This sets out a number of proposals relating in particular to high-risk AI systems, one specific definition of which is given in Annex III part 3) b) as:
\begin{quote}
"AI systems intended to be used for the purpose of assessing students in 
educational and vocational training institutions and for assessing participants in 
tests commonly required for admission to educational institutions."
\end{quote}

The proposed EC definition of an AI system seems both incomplete and flawed, but it may be safe to assume it will cover almost all forms of algorithmic processing in the final regulations. 

The consequences of such regulations, if also enacted in the UK, would be to require that operators of algorithms are able to explain how algorithms came to decisions. In general making the operation of machines intelligible to humans may not be practical for many complex algorithms. In a mixed intelligence system, this may be less of an issue since the human is making the final judgement informed by the machine rather than directed by it. However, a risk remains that the machine assessment may subliminally influence the human marker.

As we write this there is a global focus on making AI systems unbiased and fair. It is very much an open question where sources of bias come from, are they algorithmic or human or both. Very often the machine automation of human processes encodes or reflects very human biases, rather than fundamental machine biases. In the long term, it may prove easier to design machine processes that have neutral biases than to change human behaviour to be unbiased. This suggests ethically designed machine assessment could have long term benefits in producing fairer assessment outcomes. 

\begin{table*}[htb]
  \centering
  \includegraphics[width=.95\linewidth]{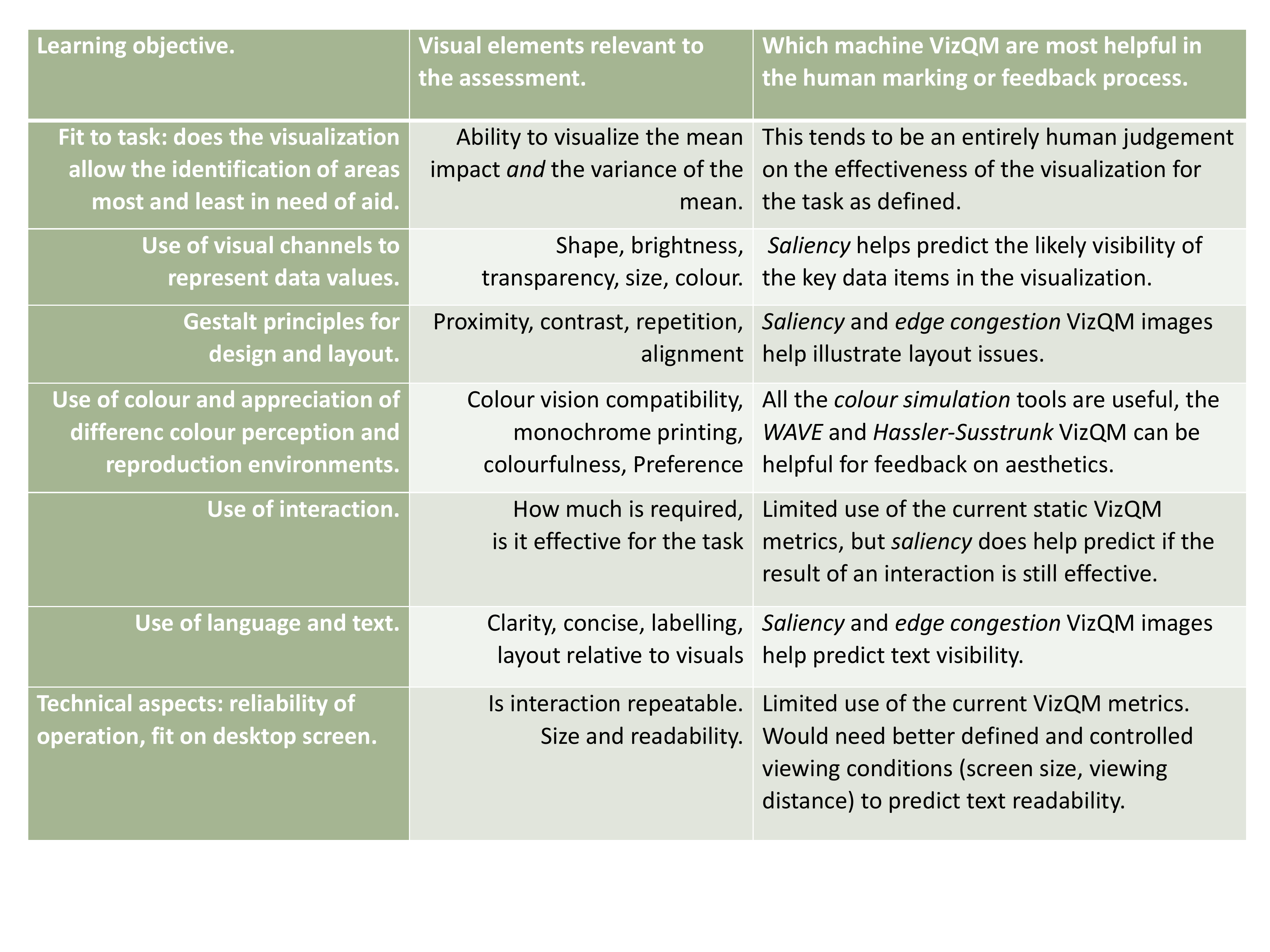}
  \caption{\label{fig:appendix}
           The relevant part of our marking scheme detailing the learning objectives, the elements that relate to these and a concluding remark on which machine assessment VizQM have seemed most helpful during the assessment process.}
\end{table*}

\section{Conclusion}
The conclusions in Table~\ref{fig:appendix} summarise which of the machine assessment methods have contributed most to the human marking and feedback. The helpfulness of the machine feedback varies with assessment task but the mixed intelligence approach certainly brought new factual information into the assessment process. We conclude by reflecting on the two objectives for automatic assessment processes that were suggested by Wilcox~\cite{Wilcox2015}.

\subsection{Does the proposed automation contribute to or detract from student learning?}

In the work here we believe, based on the positive feedback from our students, that the mixed intelligence approach helps engage students in the feedback report and encourages reflection on their work in new ways. 

The mixed intelligence approach also helps clarify some marking and feedback decisions that need to be made by the marker, for example by visualizing colour problems directly and by ranking against simple metrics from previous submissions. This provides automatically generated evidence that a marker can set their assessment and feedback against. 

\subsection{Do the benefits outweigh the costs of automation?}

The costs are noticeable, even building a system partially reusing existing code-bases, the toolset here has taken substantial time over two years to bring together and test. While it is difficult to compare quantitatively the mixed intelligence approach does now seem to be faster than the previous human-only approach. One reason for this is that the machine assessment is supporting some subjective judgements on visualization much more quickly than relying entirely on human interpretation, for example on colour and layout.

It is our plan to keep using the mixed intelligence approach and with less development overhead in future, this may mean future time savings in the marking process.
Monitoring for bias and errors will need to be a routine activity and will become easier as we gather more data on the effectiveness of the system.

\subsection{Future directions}

There is scope for studying a wider range of image informatics approaches as well as evaluating new versions of these methods such as Deep Gaze II. 
In addition recent work on verbalizing visualizations~\cite{henkin2020}, if it can be automated, could be adapted to suggest natural language machine feedback as well as the current image and metric feedback.

Tools for monitoring bias in machine assessment need further consideration, as do methods for producing balanced machine assessment feedback. One approach would be to use multiple saliency algorithms and only report saliency predictions for areas in the image where the majority of algorithms agree.

The methods we have used show promise in supporting assessors by providing new forms of evidence to underpin subjective marking processes. One wider challenge is how many visualization teachers would agree with the learning objectives themselves, a subject we are sure is open to future debate.

\section{Acknowledgements}
First, we need to thank all the students on Newcastle University Masters degrees who took the CSC8626 module. The development of the assessment process here would not have been possible without their hard work. 

In addition, we would like to acknowledge the previous work of all those who developed the code that is (re)published in the AIM project repository~\cite{Oulasvirta:2018}.

Finally, we thank the Alan Turing Institute for supporting the Newcastle Seedcorn project "Automating Data Visualization" via EPSRC EP/N510129/1 and for supporting Nick Holliman's Turing Fellowship.


\bibliographystyle{eg-alpha-doi}
\bibliography{CGVC_bib.bib}

\end{document}